\documentstyle[12pt]{article} \begin{document}

\overfullrule 0 mm
\language 0
{\it Dedicated to the 90-anniversary of Moscow State University
Professor Anatolii A.Vlasov (1908-1975)}
\vskip 1.5 cm
\centerline { \bf{STABILITY OF UNIFORM RECTILINER MOTION}}
\centerline { \bf{OF CHARGED SHELL}}
\centerline { \bf{ IN SOMMERFELD MODEL WITH
SELF-ACTION}} \vskip 0.5 cm \centerline {\bf{ Alexander A.
Vlasov}} \vskip 0.3 cm \centerline {{  High Energy and Quantum
Theory}} \centerline {{ Department of Physics}} \centerline {{ Moscow
State University}} \centerline {{  Moscow, 119899}} \centerline {{
Russia}} \vskip 0.3 cm {\it It is proved that in classical
electrodynamics the uniform rectilinear motion of charged "rigid"
sphere with self-action (Sommerfeld model) is stable.}

03.50.De
\vskip 0.3 cm

In classical electrodynamics long time ago in  Sommerfeld works [1,
see also 2] was derived the expression of self-force acting on
"nonrelativistically rigid charged sphere", i.e sphere with radius
$a$, its center moving along trajectory $\vec R(t)$, with total
charge $Q$ and charge density (in laboratory reference frame)
$$\rho(t,\vec r)={Q\over 4\pi a^2}\delta(|\vec r- \vec R|-a)$$.

In the case of
shell rectilinear motion this force has the form

$$F_{self}={Q^2 \over 4 a^2}\left[ -c \int\limits_{T^{-}}^{T^{+}} dT
{cT-2a \over L^2} + \ln {{L^{+}\over L^{-}}} + ({1\over
\beta^2}-1)\ln { {1+\beta \over 1-\beta}} -{2\over \beta} \right]
\eqno(1)$$ here $cT^{\pm}=2a \pm L^{\pm},\ \
L^{\pm}=|R(t)-R(t-T^{\pm})|,\ \ L=|R(t)-R(t-T)| $.

The total shell equation of motion then will be
$$m{d \over dt}(\gamma v)=F_{self}  \eqno(2)$$

This equation has one trivial solution - the uniform motion without
radiation:
$$R(t)=R_0+vt \eqno(3)$$
For linear analysis of stability of  this solution let us take
$$R(t)=R_0+vt+\delta(t),\ \ R(t-T)=R_0+v(t-T)+\delta(t-T),\ \
L=vT+\delta(t)-\delta(t-T), \eqno(4)$$
here $\delta(t)$ - small perturbation.

Then in linear approximation the self-force (1) reads
 $$F_{self}={Q^2 \over 4 a^2}\left[
{4\dot{\delta}\over \beta^2}\left(1-{1\over 2\beta}\ln{{(1+\beta)
\over(1-\beta)}}\right)-{2c\over v^3}\int\limits_{T^{-}}^{T^{+}}dT
{cT-2a\over T^3}\delta(t-T)\right] \eqno(5)$$
here $T^{\pm}={2a\over c(1\mp \beta)},$
and the equation of shell motion is
$$m\gamma^3\ddot{\delta}= F_{self} \eqno(6)$$
Solution of eq. (6) we take as $\delta \sim
\exp{(pt)}$.  Then the parameter $p$ must obey the
following equation [3] $$p^2={Q^2\over
a^2m\gamma^3}\left[ {p\over c\beta^2} \left(1-{1\over
2\beta}\ln{{(1+\beta) \over(1-\beta)}}\right) -{c\over
2v^3}\int\limits_{T^{-}}^{T^{+}}dT {cT-2a\over T^3}\exp{(-pT)})\right]
\eqno(7)$$

This equation one can rewrite in the form:

$$q^2 = -A^2 \int\limits_{x_{-}}^{x_{+}}dx\cdot {x-2\over x^3}\cdot
\left[ qx+ \exp{(-qx)} \right] \eqno(8)$$

here $q=pa/c,\ \ x_{\pm} ={2 \over 1\mp \beta},\ \ A^2=
(r_{cl}/a)\cdot (1/(2\gamma^3 \beta^3) ),\ \ \beta\not= 1$.

Dividing $q$ into real and imaginary parts:
$$q= \eta + i\cdot w$$
we split eq.(8) into two:
$$\eta^2 -w^2 = -A^2 \int\limits_{x_{-}}^{x_{+}}dx\cdot {x-2\over 
x^3}\cdot
\left[ \eta x+ \exp{(-\eta x)}\cos{(wx)} \right] \eqno(9a)$$
$$2\eta w = -A^2 \int\limits_{x_{-}}^{x_{+}}dx\cdot {x-2\over
x^3}\cdot \left[ xw- \exp{(-\eta x)}\sin{(wx)} \right] \eqno(9b)$$
Equations (9) are invariant under transformations $w\to-w$, so both
signs of $w$ are solutions, thus further we can take $w$ as positive
quantity.

 Integrand in (9b) one can  divide into two functions:
$$f_1={x-2\over x^3}$$ and $$f_2=\left[ xw- \exp{(-\eta x)}\sin{(wx)}
\right]$$  Intergal of $f_1$ over $x$ between limits $x_{\pm}$ has
zero value; $f_2$ - is positive function with
positive derivative for $\eta,\ w>0$ and $x>0$.  Thus for these
values of $\eta$ and $w$ the  intergal of $f_1\cdot f_2$   is
positive and the sign of R.H.S.  is negative. But this result
contradicts with the sign of L.H.S., which is positive for $\eta>0,\
\ w>0$.  Consequently the equations (9) have solutions only for
negative values of the real part $\eta$ of $q$.

 This proves that the uniform rectilinear motion of Sommerfeld charged
 sphere with $\beta<1$ is stable!

  \vskip 0.5 cm
 \centerline {\bf{REFERENCES}}

  \begin{enumerate}
\item A.Sommerfeld, Gottingen Nachrichten, 29 (1904), 363 (1904), 201
  (1905).
\item P.Pearle, in {\it Electromagnetism}, ed. D.Tepliz, Plenum, NY,
1982, p.211.
\item Alexander A.Vlasov, physics/9804001.
\end{enumerate}

 \end{document}